\documentstyle[12pt,aaspp4]{article}
\tightenlines

\def\lc{light curve}
\def\ml{microlensing} \def\mo{monitor} 
  
\def\ev{event}

\def\rd{Di\thinspace Stefano} 
\def\bl{binary lens}

\def\cc{caustic crossing}

\def\mag{magnification}

\def\rd{Di\thinspace Stefano}

\def\od{optical depth}
\def\ml{microlensing}

\def\mo{monitor}

\def\los{line of sight}
\def\ev{event}

\def\bl{blending}
\def\mage{magnification}

\def\cc{caustic crossing}

\begin{document}
\vskip -.8 true in 
\title{
On the Nature and Location of the Microlenses}   

\author{Rosanne Di\thinspace Stefano}
\affil{Harvard-Smithsonian Center for Astrophysics, Cambridge, MA 02138}

\begin{abstract}
\vskip -.3 true in 
This paper uses the caustic crossing events in the
microlens data sets to explore the nature and location of the lenses.
We conclude that the large majority of lenses, whether they are
luminous or dark, are likely to be binaries. 
Further, we demonstrate that 
blending is an important feature of all the data sets.  
An additional interpretation suggested by the data, that the caustic crossing
events along the directions to the
Magellanic Clouds
are due to lenses located in
the Clouds, implies that most of the LMC/SMC events
to date are due to lenses in the
Magellanic Clouds.
All of these conclusions can be tested.
If they are correct, a large fraction of lenses along the direction to the LMC 
 may be ordinary stellar binary systems, just as are the majority of the lenses
along the direction to the Bulge. Thus, a better understanding
of the larger-than-anticipated value derived for the Bulge 
optical depth may allow us to
better interpret the large value derived for the optical depth to the LMC.
Indeed, 
binarity and blending in the data sets
may illuminate connections among
several other puzzles:
the dearth
of binary-source \lc s, the dearth of non-caustic-crossing perturbed
binary-lens \ev s, and the dearth of obviously blended point-lens
\ev s. 
 
\end{abstract}
\vskip -0.4 true in 
\vskip -0.3 true in

\section{Lens Binarity as a Tool to Study Lens Populations}

\vskip -0.1 true in
\subsection{Overview}  
\vskip -0.1 true in
This paper illustrates what the study of an ensemble
of binary-lens \ev s can teach us about the total lens population.
\S 1.2 defines useful categories of \ev s, including \cc\ \ev s.
The heart of the paper is \S 2, where I show that even the relatively
small number of \cc\ \ev s observed to date in Baade's window 
and toward the LMC and SMC, indicate that the large majority of lenses are
likely to be binaries. 
The use of \cc\ \ev s to study blending is also elucidated; the \cc\ \ev s 
detected so far indicate that blending is ubiquitous. 
\S 3 addressees the likely location of the lenses;
in this regard, individual \cc\ \ev s have already provided valuable 
information. Indeed, analysis of the \cc s indicates that
both the LMC and SMC binary lenses are 
most likely to be located in the Magellanic Clouds and not in the 
Galactic Halo (Afonso {\it et al.}
1998; Albrow {\it et al.}
1998; Alcock {\it et al.} 1998; Udalski {\it et al.} 1998;
Bennett {\it et al.} 1996; Alcock {\it et al.} 1997a).
Coupling these prior analyses with the results of \S 2 leads to
the conclusion that most of the lenses detected to date along directions 
to the Magellanic Clouds are actually in the Magellanic Clouds.  
The main results of this paper, that most lenses detected to date
are binaries, that a significant portion of
the lenses detected toward the Magellanic Cloud are part of the
Magellanic Clouds, and that blending is ubiquitous,
can be tested.  
The tests may provide answers to several puzzles presented by
the data sets. The puzzles are outlined and discussed in \S 4.
The fourth puzzle we discuss is of central importance
to learning about dark matter through the microlensing observations.
That is: why are the measured values of the optical depth   
higher than predicted for both the Bulge and the LMC?
If the LMC lenses are ordinary stellar systems, as those 
along the direction to the Bulge are thought to be, the reasons
for the unexpectedly high values may be essentially, or at least partly,
 the same.
In $\S 5$ we ask: 
can blending and binarity themselves lead to overestimates of $\tau?$
We sketch the
relevant considerations and point out that further work, 
largely on the part of the
observing teams, is necessary to test the hypothesis that 
the combination of blending and binarity can cause significant mis-estimates of 
$\tau$ when these effects are unrecognized or only partially taken into 
account.         

\vskip -0.1 true in
\vskip -0.1 true in
\vskip -0.1 true in
\vskip -0.05 true in
\subsection{Categories of Binary Lens Light Curves}  

\vskip -0.1 true in
\noindent {\bf (1) Point-lens-like light curves:}\ The vast majority of \ev s
in which a binary system serves as a lens produce light curves 
indistinguishable
from point-lens \lc s. 
This is because: 
(a)  
values of $a,$ the orbital separation, are likely to be distributed 
over a wide range; thus, many binaries are either too close or too wide
to yield \lc s with measurable evidence of lens binarity, and 
(b) even in the range in which the influence of lens binarity
is most likely to be evident ($0.1\, R_E < a <  3\, R_E)$, 
a large fraction of
\lc s can be practically indistinguishable from point-lens \lc s
(\rd\ \& Perna 1997). 

\noindent {\bf (2) Caustic Crossings:}\ \ Caustic crossing \lc s are
distinctive: they exhibit wall-like structures with measured values of the
peak magnification ${\cal O}(10).$ 
Neither the presence of
the perturbation nor, in most cases,
its nature as a binary-lens phenomenon are easily 
obscured by other astronomical effects, such as the
blending of light from the lensed star with light from other
sources located in the seeing disk. 
 
\noindent {\bf (3) Other Significantly Perturbed Light Curves:}\ 
Close approaches to a caustic 
can lead to striking perturbations; many such \ev s would be included
among the ``strong" binary events discussed by Mao \& Paczy\'nski (1991).
Other significant perturbations would  
be of a more ``gentle" nature. 
Both strong and gentle perturbations can be systematically 
identified, classified, and studied
(\rd\ \& Perna 1997).

\noindent On a theoretical level, only members of the second category are well-defined.
To distinguish between members of
categories (1) and (3), on the other hand, 
specific criteria (such as the difference between binary-lens
and point-lens fits) must be chosen.
Observationally, 
membership among the three categories can be blurred by issues such
as the frequency of sampling, photometric uncertainties, the presence of
additional light from other (unlensed) stars, and the methods
chosen to analyze the data. 

Because \cc\ \ev s are the only types of binary-lens \ev s that have
been reliably classified so far, it makes sense to concentrate on 
what we can learn from them. Fortunately, 
caustic crossing \ev s provide us with the clearest view of important
elements of the data set, including the fraction of all events likely to
be due to binary lenses and the role played by blending throughout the 
data set.

\vskip -.3 true in 
\vskip -.2 true in 
\section{What We Learn from Caustic Crossings}
\vskip -0.1 true in

We will conclude that it is not presently
possible to falsify the hypothesis that {\it all} of the lenses
observed to date are binaries.  Sketched below, the arguments that lead to this 
conclusion also imply that blending is playing an important role in 
shaping the data sets.

\vskip -.1 true in 
\vskip -.1 true in 
\vskip -0.1 true in
\vskip -0.1 true in
\subsection{The Observed Relative Frequency of Caustic Crossing Events}
\vskip -0.1 true in
The published LMC data contain $1$ \cc\ \ev\ and $7$  point-lens-like
\ev s (Alcock {\it et al.} 1997a~\footnote{At least  
one perturbation that passes all of the cuts 
for a point-lens \ev\ exhibits some anomalies that have caused it to be
excluded from the detailed statistical analysis of the LMC \ev s.  
};    
Bennett {\it et al.} 1996).
Thus, \cc\ \ev s constitute slightly more than $10\%$ of the LMC \ev s. 
This is 
consistent with the fraction obtained for the Bulge data 
set. 
(See, e.g., Udalski {\it et al.}\ 1994; Alard, Mao, \& Guibert 1995;
Alcock {\it et al.}\ 1997b~\footnote{For both
the OGLE and DUO teams, \cc\ \ev s constitute $\sim 10\%$
of all Bulge \ev s 
(Udalski {\it et al.}\ 1994; Alard, Mao, \& Guibert 1995). 
The fraction of the MACHO team's published
Bulge \ev s exhibiting \cc s is considerably smaller 
(Alcock {\it et al.} 1997b). With regard to the MACHO results we note 
that (1) the algorithms used to identify the \ev s published so far were
designed to exclude perturbations not of the point-lens form, and (2) the 
present observing strategy of calling an ``alert" which institutes frequent 
high-precision monitoring of many \lc s,
will increase the rate of binary-lens discoveries. As a matter of
fact there are $\sim 5$ Bulge \cc\ \ev s listed on the alert web pages covering
1996-1998.}).  
For the Bulge, and especially for the LMC, the total number of 
published \cc\ \ev s is
still small;
it is nevertheless worth considering the conclusions we can draw if
\cc\ \ev s continue to constitute at least a few percent of all \ev s.  
The SMC has not been observed for as many star-years (the product
of the number of stars monitored and the time during which they are
monitored) as either the 
LMC or the Bulge. 
The published SMC data consist of
 $1$ \cc\ \ev\ (Afonso {\it et al.} 1998; Albrow {\it et al.} 1998;
Alcock {\it et al.} 1998; Udalski {\it et al.} 1998;
Rhie {\it et al.} 1998) and $1$ point-lens-like
\ev\ (Alcock {\it et al.} 1997c; 
Palanque-Delabrouille {\it et al.} 1997; Udalski {\it et al.} 1997; Sahu 1998).
 
\vskip -.1 true in 
\vskip -.2 true in 
\vskip -.3 true in 
\subsection{The Predicted Relative Frequency of Caustic Crossing Events}

\vskip -.1 true in 
Let $N_{cc}/N_{tot}$ represent the fraction of all \ev s exhibiting \cc s. 
The larger the predicted value  
of $N_{cc}/N_{tot}$,  
 the more likely it is
that some of the point-lens-like \ev s already observed
are actually due to lensing by a point mass rather than lensing by
a binary. 
To determine a reasonable upper limit for $N_{cc}/N_{tot},$ 
I carried out a set of simulations in which all lenses were binaries
with orbital separations between $0.1\, R_E$ and $1.6\, R_E.$
The binaries were also characterized by a value of $q,$ the ratio of the 
masses of the less massive to more massive component. 
I considered 
$20$ values of $q$ (starting with $q=0.05$ and
proceeding by intervals of $0.05$ to $q=1.0$), and $5$ values of $a,$
the semi-major axis 
($a=[.1]\, 2^{i-1} R_E,$ with $i$ ranging from $1$ to $5$), and computed the 
rates of events in each of the $3$ categories sketched in \S 1. 
The bottom-most (thick) curves in Figure 1 show the results. 
Averaging over the values of $a$ considered, 
$N_{cc}/N_{tot}$ is maximized for values of $q$
near unity. In fact, $N_{cc}/N_{tot}$ is approximately equal to
$0.15$ for $q > 0.4.$   
Thus, even if all lenses are binaries with $a$ and $q$ in the ranges that maximize the
size of caustic structures, each \cc\ \ev\ should be accompanied by
$6-7$ additional \ev s. $2-3$ of these additional \ev s are likely to be
significantly perturbed from the standard point-lens form; $3-5$ \ev s
will, with the \lc\ sampling used in the simulations,
 be indistinguishable from point-lens \ev s. 

Of course, even if all lenses are binaries, 
\ev s in which the orbital separation between the lens components
is in the range considered in the simulations will 
form only a fraction ${\cal F}_{ev} < 1$ of all \ev s. For populations such as
those in our disk, it seems reasonable to estimate ${\cal F}_{ev} \sim 1/3,$ although
this number can be smaller or somewhat larger, depending on the distributions
of values of the mass ratio and orbital separation. 
Thus, each \cc\ \ev\ should be accompanied by 
$\sim 15$ or more point-lens-like \ev s.  Even more point-lens-like \ev s
are expected when the lens population is not entirely composed of
binaries.
\footnote{It is interesting to note that the lenses used in the simulation
are most likely to be representative of a true lens population 
located in the core of a globular cluster, where (1) the fraction of stars that
are in binaries may be large due to the effects of mass segregation and
stellar capture processes, and (2) orbital separations are expected to be
set by ambient translational velocities on the order of $10$ km/s,
and to therefore be in the range of several au, fairly
close to reasonable values of $R_E.$}

At first glance, this would indicate that, relative to theoretically-derived
expectations,  
\cc\ \ev s are overrepresented in the data sets.
When, however, we take into account that
the fields being monitored are crowded, and that some of the lens systems may 
themselves be luminous, 
it is clear that blending of light from different sources
along the line of sight to an \ev\ may be important. 
In \S 2.4 we test the influence of blending 
on the value of $N_{cc}/N_{tot}.$
In \S 2.3 we show that \cc\ \ev s  provide a direct measure
of the role blending plays in \ml\ observations.

\vskip -.3 true in 
\vskip -.1 true in 
\subsection{What Caustic Crossing Events Can Teach Us About Blending}

\vskip -.15 true in 
Let $A_{lim}$ be the minimum value of the peak \mage\ required in order
to reliably conclude that an observed \lc\ perturbation is due to \ml .
If $f$ is the fraction of the baseline flux provided by a lensed star,
then its true \mage\ must be greater than
\begin{equation}
A = 1 + \Bigg[{{A_{lim}-1}\over{f}}\Bigg], 
\end{equation}
in order for the \ev\ to be correctly identified.
Given this, \cc\ \ev s provide a clear advantage for the detection
of heavily blended (i.e., small $f$) \ev s. 
First, the \mage\ during the \cc\ can be
so large that only the most heavily blended \ev s would not be detected during 
such a crossing. Second, because there are always 2 crossings
(which should generally be resolvable), there are 2 chances to catch
even a heavily blended \ev . Third, the minimum magnification
after the first \cc\ during the interval until the second crossing
is 3 (Witt \& Mao 1995); this, together with the large \cc\  \mage, means that,
even with $A_{lim}$ as high as $1.34,$ \cc s should allow for 
extremely efficient sampling of \ev s with $f > 0.17.$
In fact, for $f > 0.17,$ the distribution of $f$ values in \cc\ \ev s should mirror
the true
distribution of values in all lensing \ev s. 

The fits to all of the published \cc\ \ev s require
significant blending. With $f$ defined as the fraction of the baseline
flux due to the lensed star: $f_I = 0.56$ for the OGLE 7 \ev\
(Udalski {\it et al.} 1994);
$f_B = 0.73, f_R = 0.70$ for the DUO team's \cc\ \ev\ 
(Alard, Mao, \& Guibert 1995);   
$f_B = 0.17, f_R = 0.26$ for the LMC \cc\ \ev (Bennett {\it et al.} 1996); 
$f_V = 0.57, f_R = 0.49$ for the SMC \cc\ \ev\ (Rhie {\it et al.} 1998;
but see also the compendium of fits included therein).\footnote{
Note that, among the papers quoted,
 the subscripts indicating color do no generally refer to the 
same wavelength ranges.}  
Thus, although it is too early to extract a distribution of $f$ values,
\cc\ \ev s give 
a clear indication that blending is important along each direction presently
being monitored. This is consistent with what we have learned about
blending in other ways (Alard 1998). The apparent importance of blending throughout the 
data set means that its effects should be included in our estimates
of $N_{cc}/N_{tot}.$   

\vskip -.4 true in
\subsection{Blending and the Value of $N_{cc}/N_{tot}$}

\vskip -.15 true in 
 The fraction
of the baseline flux emanating from the lensed source, $f,$ was 
smaller than unity in $5$ simulations: $f = 0.1 + 0.2\, (j-1);\, 
j=1,5.$ The results are summarized in Figure 1. 
When blending is moderate ($f > 0.6$), the results do not differ
significantly from the unblended case.
As blending becomes more important,  
 the fraction of \ev s in which there are caustic crossings rises. 
The results for the real data sets
 depend on the distribution of values of $f;$ there are consistent models
with moderate blending in which $1$ \cc\ \ev \ is 
accompanied by ${\cal O}(1)$ member of category $3,$
and $\sim 10$ point-lens \ev s.

\vskip -.2 true in 
\vskip -.1 true in         
\subsection{Conclusions}
\vskip -.15 true in 

The work described above indicates that each \cc\ \ev\ should be accompanied
by a number of other \ev s, and that the number of other \ev s can be
on the order of $10$ for moderate blending. 
Occam's razor, appled to the observations, thus dictates that these
additional \ev s form the major component of the \ev s we have actually
already seen.   
The majority of lenses giving rise to the \ml\ \ev s observed so far
are almost certainly binaries. 

\vskip -.2 true in         
\vskip -.1 true in         

\section{Where Are the Lenses?}  
 
Information gleaned from studying the \cc s in both the SMC and LMC data
indicate that the lenses in these specific \ev s are
most likely located in the Magellanic Clouds themselves (Afonso {\it et al.}
1998; Albrow {\it et al.} 
1998; Alcock {\it et al.} 1998; Udalski {\it et al.} 1998;
Bennett {\it et al.} 1996; Alcock {\it et al.} 1997a).
 Should 
these conclusions be bolstered by future \ev s, 
and if it is true that the lenses giving rise to point-lens-like \ev s
are drawn from the same population as the
lenses giving rise to the \cc\ \ev s, 
we must conclude
that it is likely that {\it all} of the \ev s are due to lenses
in the Magellanic Clouds rather than in the Halo. Conversely, if
the location of the lenses implicated in the \cc\ \ev s is established
to be in the Halo, then we would have discovered that Galactic MACHOs 
tend to be binary systems.

\noindent {\bf Tests:\ } At present it seems most likely that  
the lenses are in the Magellanic Clouds. 
This conclusion    
 carries a number of testable consequences. First,
many LMC lenses will be luminous, and the degeneracy inherent in the 
point-lens \lc\
is likely to be broken in some cases (Udalski {\it et al.} 1994; \rd\ \& Esin 1995;
Kamionkowski 1995); this can 
determine the location of
the lenses associated with even some point-like \ev s.
Second, if the lenses are in the LMC and are luminous binaries,
then there are opportunities for consistency checks
between the known binary population (which becomes better known
through the frequent observations carried out by the observing teams),
the population of detected binary lenses and, eventually, binary sources.
We note that there is direct evidence that
the single SMC \ev\ without the \cc\ is located in the SMC. Furthermore, there
are indirect arguments that the majority of the Bulge lenses are in the
Bulge (see, e.g., Kiraga \& Paczy\'nski 1994).  
 The open question, and the one with the most important
implications vis-a-vis the composition of the Halo, is the location of
the LMC lenses.  Additional \cc , and/or 
members of category $3$, 
will therefore play a crucial role in helping us to understand the nature
and location of the LMC lenses.

\vskip -.2 true in
\vskip -.1 true in
\section{Tests and Puzzles} 

\vskip -.15 true in 
\subsection{Tests}

\vskip -.15 true in 
The conclusions of the preceding two sections are testable. Some tests were
discussed in \S 3. Here we provide an overview, focusing on the connection
between the phenomena of blending and binarity and some puzzles posed by
the existing data. First, the overview. If the lenses are indeed
binaries, 
we should continue to discover \cc\ \lc s. Even if all or most
lenses are binaries, the average
fractional rate could be
significantly lower than that observed so far, and will depend on the
distribution of values of $q$ and $a$ among the lens population.
A second important test is that perturbed \ev s that are not 
\cc\ \ev s should be observed. Finally, 
if the \ev s are primarily due to lenses in a single binary population,
several consistency checks are possible. One example: the \cc\ \ev s
can, as a group be used to extract the probability distribution for
$q;$ this distribution should be consistent with the results of a
parallel analysis using only members of category $3$. As a larger
ensemble of events is collected, the ratio $N_{cc}/N_3$ can be
measured as a function of $f,$ $q,$ and $a;$ the results can be
compared with theory. 

\vskip -.1 true in 
\vskip -.1 true in 
\subsection{Puzzles}

\vskip -.15 true in 
\noindent {\bf 1. If blending is ubiquitous,  
why do so few point-lens-like \ev s exhibit evidence of \bl ?\ }
Evidence of blending
in the \lc\ is most noticably encoded in the discrepancy between
the peak and baseline fluxes, relative to a point-lens model 
(\rd\ \& Esin 1995). It is easier to detect these discrepancies
when the peak \mage\ is large. 
In fact, depending on the photometry and frequency of sampling, the distance of
closest approach to a point lens needs to be $\sim 0.2-0.3$ in
order for blending to be detected using the \lc\ alone (\rd\ \& Esin 1995,
Wozniak \& Paczy\'nski 1997).  
When the lens is a binary, 
the enhancement in peak \mage\ associated with \cc s 
can provide an ideal opportunity
 to look for blending. Indeed, during the first few years of
observations, \cc\ \ev s were the only known blended \ev s, and
to date, all \cc\ \ev s are blended.

\noindent 
The importance of the role of blending can be tested in many ways.
(See, e.g., \rd\ \& Esin 1995;
Kamionkowski 1995; Buchalter, Kamionkowski, \& Rich 1996;
Wozniak \& Paczy\'nski 1997; Goldberg, Wozniak, \& Paczy\'nski 1997;
Goldberg 1998; Goldberg \& Wozniak 1998; Han 1998)
Indeed, by using using the image subtraction method to re-examine
the OGLE team's Bulge data, Alard finds that the fraction of \lc s
exhibiting signs of blending is consistent with the fraction
predicted by \rd\ \& Esin (1995) and Wozniak \& Paczy\'nski (1997).
Future spectroscopic and astrometric studies can
provide complementary information. Thus,  
detailed answers to this first puzzle seems well within reach.
The next two puzzles may, at this point in time, be compared to Conan Doyle's 
dog that didn't bark in the night.

\noindent {\bf 2. and 3. Where are the binary source events?
\   Where are the members of category $3$?\ }
In this paper I argue that the presence of \cc\ \ev s in the data sets implies
the presence of members of category 3. Griest \& Hu (1992) have argued 
that $2-5\%$ of \ml\ \lc s should exhibit significant perturbations
associated with source binarity. An independent analysis yields 
comparable results (Han \& Jeong 1998), even after 
taking into account degeneracies inherent in 
binary-source \lc s (Dominik 1998; Han \& Jeong 1998). 
Unfortunately,  no perturbed binary source \lc s are
in the published literature, which is also free of members of category  
$3$.~\footnote{It has been established, however, that in SMC-1 a binary source 
was magnified; the discovery of the binary nature of the source was based
on residuals in the baseline flux. 
}
Some of the missing highly perturbed \lc s must be in the data set,
but are not yet identified as microlens candidates. 
~\footnote{It
seems to be the case that the Bulge data set
contains a number of such \lc s (Axelrod 1997).}
In fact, when I carried out a set of simulations like those described
in \S 2, but
attempting to mimic the MACHO team's observing pattern (frequency, including
gaps, and photometric sensitivity) and algorithmic cuts to eliminate
\ev s not deemed to be good candidates for \ml , 
the large majority of strongly perturbed binary-lens
\ev s did not pass the cuts.
It is also important to note 
that many events with physical parameters normally associated with
serious perturbations are likely published among 
point-lens/point-source \ev s; the
\lc s show no obvious perturbations
because of the blending of light from the source star or stars
with light from other stars, possibly including the lens
system, along the \los . 
Blending 
washes out distinctive non-singular features  of the perturbed \lc s.  
If, for example,
there is a secondary peak that rises $\Delta A$ above the curve
associated with a point-lens fit, blending reduced the difference to
$f\, \Delta A$. 
 Indeed, the second and fourth columns of Figure 2 show that when 
blending is severe ($f=0.1$), then the \ev s which are well-fit by point-lens
models are primarily drawn from \ev s that were in categories $2$ and $3$
when there was no blending ($f=1$). This is because 
the caustic structures enhance the peak \mage\ of a significant
fraction of the \ev s in categories $2$ and $3$, so they survive
as detectable \ev s, even when there is severe blending, although
signatures of their distinctive nature may be washed out.
Because binary-source \ev s do not benefit from such local increases in \mage ,
they are more likely to be classified as Paczy\'nski \lc s when
blending is moderate, and more likely to be missed altogether when
blending is severe.  This distinction between binary-lens and binary-source \ev s
should have observable consequences. For example, if most source and lens 
systems are binaries, highly perturbed \lc s discovered by additional thorough
searches of the data or by the careful observations of the follow-up teams
are more likely to be binary-lens \ev s. In addition, perhaps $1\%$ of all
\lc s should display truly unusual features, due to the lensing of
a binary source by a binary.  

\noindent {\bf 4. Why Are the Event Rates and the Measured Values of the Optical
Depth So High?\ }
Along the direction to the LMC the measured value of the \od\ is
$2.1^{+1.3}_{-0.8} \times 10^{-7},$ 
while estimates of the component due to stars is smaller than 
$0.5 \times 10^{-7}$ (Alcock {\it et al.} 1997a, Renault 
{\it et al.} 1997; see also Bennett 1998). It is
tempting to ascribe the discrepancy to the presence of dark
matter in the Halo (but see Sahu 1994a, 1994b; Wu 1994).
Proof that the
lenses are in the LMC may be welcome, 
since Halo models of a MACHO population of masses in the range
indicated by the \ml\ results have proved difficult to make
compatible with other observations that should be have found
evidence of this population or its precursor.  (See, e.g.,
Graff \& Freese 1996; Adams \& Laughlin 1996; Chabrier, Segretain, \& Mera 1996;
Gibson \& Mould 1997; Fields, Mathews, \& Schramm 1997.) 
Problems
with the LMC-lens interpretation have, however, also been
noted; the most serious of these
is that the virial theorem provides an independent measure the optical
depth (Gould 1995), and measurements of the speeds of some tracer stars
indicate that velocities are too small to be consistent with a 
large \od\ (Cowley \& Hartwick 1991). It is of course possible that    
additional kinematic studies of LMC populations will indicate that
the \od\ of LMC lenses is indeed consistent with the present analysis of the
\ml\ data. It is also worth considering whether the published \ml\ results
are too high. A comparison with the Bulge, where the majority of
lenses are stellar systems,  may be instructive. 
Indeed, most indications are that the
\ml -measured value of the Bulge optical depth is also higher
than predictions based on other types of observations.  
The OGLE result (based on
$9$ lenses), and the MACHO result (based on $45$ lenses) were
$3.3^{+1.2}_{-1.2} \times 10^{-6},$ and $3.9^{+1.8}_{-1.2} \times 10^{-6},$
respectively, while predictions were in the range $0.5-1.0 \times 10^{-6}.$
While changes in how the Galaxy is modeled can apparently make the
predictions come into marginal agreement with the observations 
(see, e.g., Alcock {\it et al.} 1997b),
it is far from clear that the problem is solved. In fact,
using COBE data,
Bissantz {\it et al.} 1997 derive values
of $0.83-0.89 \times 10^{-6}$ for main-sequence stars, and
$1.2-1.3 \times 10^{-6}$ for red-clump giants. Presumably the
discrepancies between predictions and observations will be resolved by
an explanation based on features of the stellar population thought to
provide most of the lenses. Thus, whatever the solution  
for the Bulge, it may take us a long way toward the solution for the LMC.
Can the combination of binarity and
blending provide an important part of this solution? 

\vskip -.1 true in 
\vskip -.1 true in 
\vskip -.2 true in 
\section{Binarity and the Optical Depth}

\vskip -.15 true in 

Blending and binarity both 
need to be systematically
incorporated into the data analysis. 
Their influence on \ev\ detectability can be 
understood geometrically. 
Consider \ev s with $A>>1.$ 
When the lens is a point mass, 
the high-\mag\ region is a disk of radius $1/A$ centered on the lens.
For blended \ev s, the high-\mage\ region is a disk
of smaller radius, $f/(A-1) \sim f/A.$ 
\footnote{This can be derived by considering the 
appropriate series expansion of the 
the blended Einstein radius, 
$R_{E,bl},$ which is smaller than $R_E$ by a factor
$\sqrt{2}\sqrt{{{(A_{min}-1)+f}\over{\sqrt{(A_{min}-1)^2
+ 2 (A_{min}-1)\ f}}}-1}$ (\rd\ \& Esin 1995).
}  
When the lens is a binary, the high-\mage\ regions are closed 
ribbons which trace the caustic structures; there can be several
disconnected, closed high-\mage\ ribbons. Although the width of each ribbon    
is smaller for larger values of $A,$ the linear size of the region enclosed
by the ribbons 
is not much affected by increasing values of $A$, and is generally
significantly larger than $2/A,$ even when there is blending.
Thus, lens binarity allows us to  
more often reach deeply into the 
luminosity function of potential source stars.
How we interpret the additional events can influence 
observation-based estimates of $\tau$.

\vskip -.1 true in 
\vskip -.1 true in 
\vskip -.2 true in 
\subsection{Interpretation of the Data}

\vskip -.15 true in 
The \od\ has been derived from the data by using the simple relationship
\begin{equation}
\tau=
{{\pi}\over{4\, N_{obs} T_{obs}}}\, \sum_{l=1}^L{{\hat t_{E,l}}\over{\xi_l}}.
\end{equation}
$N_{obs}$ is the number of stars monitored and $T_{obs}$ is
the time during which the stars were monitored;
$l$ labels the \lc s;
$\hat t_{E,l}$ 
is the time it
takes for the source star to traverse an Einstein diameter; 
and $\xi_l$
is the efficiency for detecting a point-lens \ev\ with the duration
observed for event $l.$

Blending and binarity can introduce
non-trivial complications into the process of using this equation to
compute $\tau.$
First, {\it all} perturbations from baseline 
not clearly associated with known types of
stellar variability must be considered as microlens candidates. 
Second, for each candidate  
\lc , the full complement of acceptable fits 
which include the
effects of binarity and blending must be found;
this allows significantly blended point-lens \ev s,
and most members of
categories $2$ and $3$ to be included in the sum.
\rd\ \& Perna (1997) have shown that (1) although 
binary-lens \lc s are diverse, they are nevertheless distinctive
enough that
most non-lensing perturbations
can be eliminated; and (2)
most \lc s that can be described by binary-lens
models can be well fit by several sets of physical parameters,
$\{{\cal M}_i\}$, and that
all acceptable fits can be found in a systematic
way. The technique they developed can readily accommodate blending
(and other effects, such as source binarity) as well.
 Each fit provides a value of $\hat t_{E,l};$ thus, 
an important effect of the degeneracies is 
to increase the uncertainties
in estimates of $\tau.$ 
 In addition, the value $\xi_l$
may depend on the true nature of the underlying \ev , 
e.g., on the size of high-\mage\ regions, in addition
to the size of the Einstein ring.
Finally, the number of
source stars that can be detectably lensed must be carefully computed.
\footnote{An example is provided by blending (even when point sources
are lensed by point masses). In this case, the sum can contain several terms
per line of sight. Yet, the estimates of the value of individual terms are smaller
when there is unrecognized blending (see footnote 7), and in fact are small 
enough to fairly well balance the influence of the additional
terms.     
The arguments above indicate, however, that this is not necessarily the case
when the lenses are binaries. Instead, this question needs to be considered as 
a function of the true distributions of $f$ values and binary properties
to determine if the factor outside the sum (essentially a normalization
factor) requires alteration.}   
In fact, if blending and binarity play important roles, this single
consideration can significantly
decrease estimates of $\tau$ based on computations that did not fully consider
binarity and blending in concert.

The discussion above makes it clear that a good deal of further work
is required to quantify the influence of
blending and binarity on estimates of the optical depth. It is, however,
useful to note that: (1) 
Estimates that do not take blending and binarity
into account are likely subject to larger uncertainties than those presently
quoted.
(2) Measurements good to
within a few tens of percent will 
require that blending and binarity be systematically
incorporated into the analysis. 
(3) The total effect may be larger, but its sign and size can only be 
determined by the observing teams; 
this is because  
detailed information about blending in the fields studied,
 and the details of the analysis of data play
such important roles.

In summary, developing an integrated approach to the data analysis
that incorporates source binarity, lens binarity, and blending,
is necessary if we are to fully understand the results of the 
microlensing observations. This paper begins to demonstrate, 
however, that the effort is well worthwhile: apparently
anomalous events can, both individually and collectively, 
 encode more information than the 
standard point-source/point-lens \ev s.

\vskip .2 true in 

\centerline{{\bf Acknowledgments}} 
I would like to thank C.S. Kochanek and K.Z. Stanek for helpful comments.  
This work was supported in part by NSF 
AST-9619516, and by funding from AXAF.

\vskip .2 true in

\vskip -.2 true in 
\vskip -.2 true in 
\vskip -.2 true in

\begin{figure}
\plotone{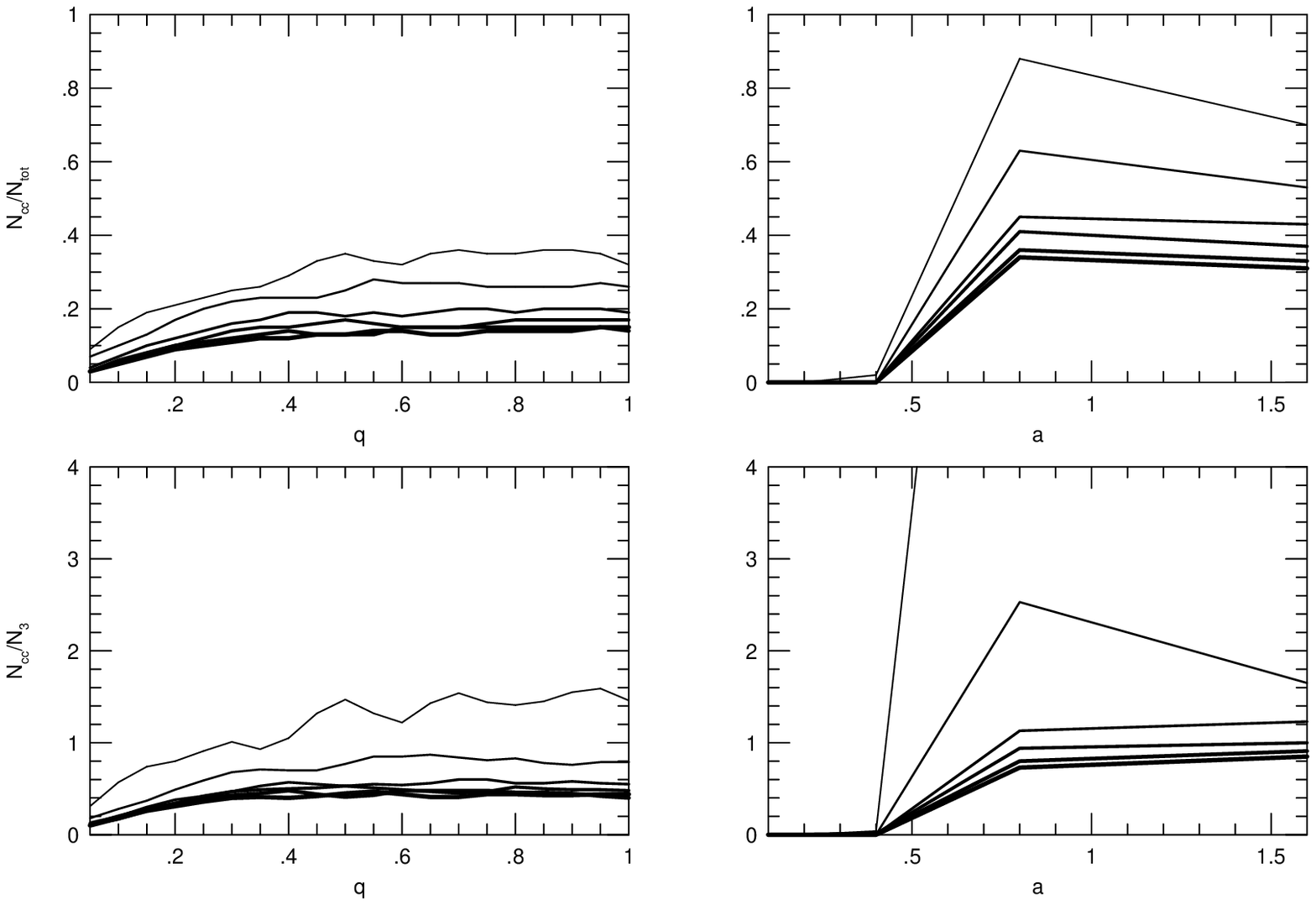}
\vspace{-3.0 true in}
\caption{
$N_{cc}/N_{tot}$ is the fraction of all detected \ev s in which there was a 
\cc . The upper left and right panels show $N_{cc}/N_{tot}$ as a function
of $q$ and $a,$ respectively. 
(An \ev\ was called ``detectable" if $A_{peak}>1.34$, and at least
10 observations were made while the source was within $1.5\, R_E$
of the lens)
In each panel, the thickest (bottom-most)
curve corresponds to the case of no blending; proceeding generally upward
to progressively thinner lines, $f = 0.9, 0.7, 0.5, 0.3, 0.1$. Note that the
differences between $f=0.7-1$, and even $f=0.5,$ are small. For smaller 
values of $f,$ the role of \cc\ \ev s becomes relatively more important.
For any given value of $f,$ \cc s are most important for moderate to
high values of $q$ and for $a$ near unity. Note that the trend seen here as
$a$ increases clearly indicates that \cc s continue to play a role,
even for larger values of $a.$ 
The lower left and right panels show the ratio, $N_{cc}/N_3,$ of \cc\ to
other perturbed \ev s. It is clear that, unless blending is extreme,
perturbed \ev s with no \cc s dominate.
}
\end{figure}
 
\begin{figure}
\vspace{-1 true in}
\plotone{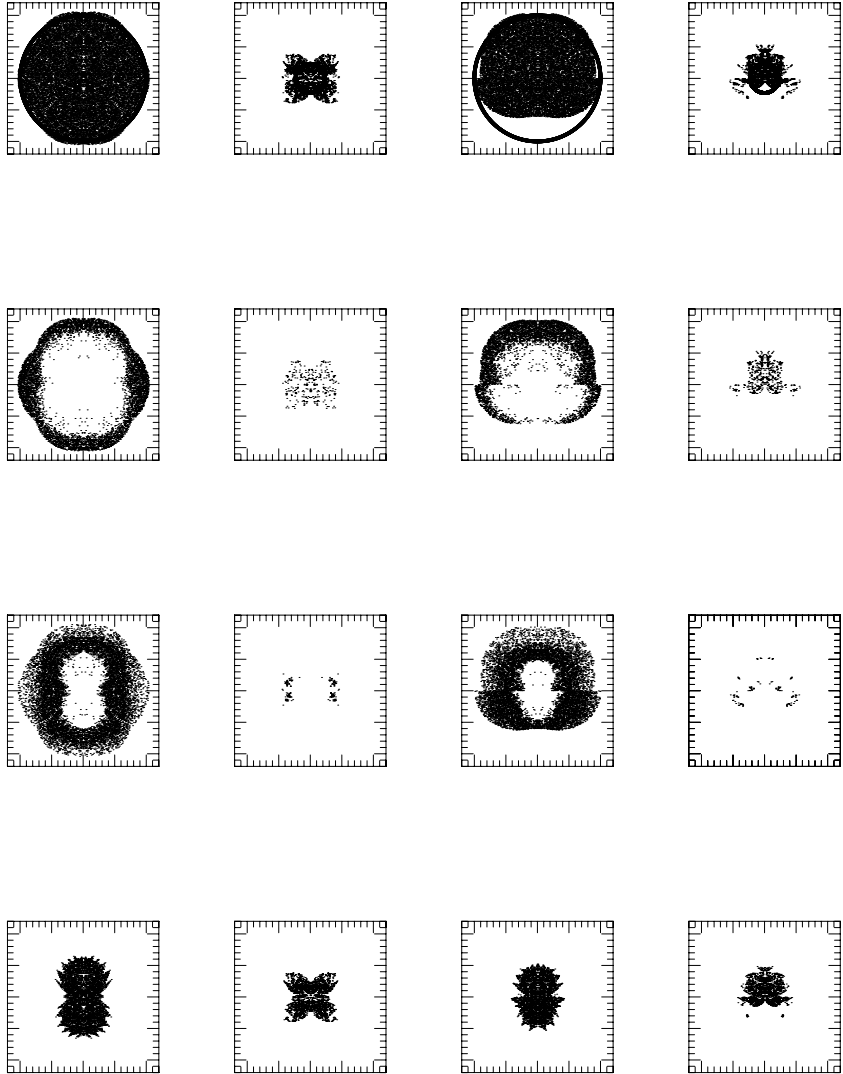}
\vspace{-.8 true in}
\caption{
Each point represents a \ml\ \ev ; the position of the point is the 
position of the closest approach to the center of mass of the 
binary lens. The horizontal and vertical axes represent the $x-$ and $y-$ axes
of the lens plane. 
{\bf Top row:}\ All measurable events ($A_{peak}>1.34$, and at least
10 observations were made while the source was within $1.5\, R_E$
of the lens).   
{\bf Second row:}\ All measurable events with acceptable point-lens fits.  
{\bf Third row:} \ All measurable events not well-fit by a  point-lens model
and in which there is {\it no} caustic crossing.  
{\bf Bottom row:}\ All measurable events not well-fit by a  point-lens model
and in which there is a caustic crossing.  
The binary depicted in the two left-most columns has $q = 1.0$, $a = 1.0$. 
The binary depicted in the two right-most columns has $q = 0.25$, $a = 1.0.$
In each set of two columns, there is no blending in the left column, and
severe blending ($f=0.1$) in the right column.  
}
\end{figure}

\end{document}